\documentclass[11pt,draftcls,onecolumn]{IEEEtran}
\usepackage{amsmath,amssymb,eucal,graphicx}
\usepackage{epsfig}
\usepackage{exscale}
\usepackage{latexsym}
\usepackage{verbatim}
\usepackage{amsfonts}
\usepackage{graphicx}%
\usepackage{enumerate}
\usepackage[usenames]{color}
\usepackage[normalem]{ulem}

\newfont{\bbb}{msbm10 scaled 500}

\newfont{\bb}{msbm10 scaled 1100}
\newcommand{\CC}{\mbox{\bb C}}
\newcommand{\RR}{\mbox{\bb R}}

\newcommand{\EE}{\mbox{\bb E}}

\newcommand{\av}{{\bf a}}

\newcommand{\gv}{{\bf g}}
\newcommand{\hv}{{\bf h}}

\newcommand{\pv}{{\bf p}}
\newcommand{\qv}{{\bf q}}

\newcommand{\uv}{{\bf u}}
\newcommand{\wv}{{\bf w}}

\newcommand{\xv}{{\bf x}}
\newcommand{\yv}{{\bf y}}
\newcommand{\zv}{{\bf z}}
\newcommand{\zerov}{{\bf 0}}
\newcommand{\onev}{{\bf 1}}

\newcommand{\Am}{{\bf A}}

\newcommand{\Gm}{{\bf G}}
\newcommand{\Hm}{{\bf H}}
\newcommand{\Id}{{\bf I}}

\newcommand{\Pm}{{\bf P}}

\newcommand{\Rm}{{\bf R}}

\newcommand{\Cc}{{\cal C}}

\newcommand{\Fc}{{\cal F}}

\newcommand{\Nc}{{\cal N}}

\newcommand{\Pc}{{\cal P}}
\newcommand{\Qc}{{\cal Q}}
\newcommand{\Rc}{{\cal R}}
\newcommand{\Sc}{{\cal S}}

\newcommand{\Uc}{{\cal U}}

\newcommand{\gammav}{\hbox{\boldmath$\gamma$}}

\newcommand{\lambdav}{\hbox{\boldmath$\lambda$}}

\newcommand{\thetav}{\hbox{\boldmath$\theta$}}

\newcommand{\sigmav}{\hbox{\boldmath$\sigma$}}

\newcommand{\Deltam}{\hbox{\boldmath$\Delta$}}

\newcommand{\diag}{{\hbox{diag}}}

\newcommand{\trace}{{\hbox{tr}}}
\newcommand{\rank}{{\hbox{rank}}}

\renewcommand{\arg}{{\hbox{arg}}}

\newcommand{\eqdef}{\stackrel{\Delta}{=}}

\newtheorem{theorem}{Theorem}
\newtheorem{definition}{Definition}
\newtheorem{lemma}{Lemma}

\newtheorem{remark}{\indent \bf Remark}[section]

\def\BibTeX{{\rm B\kern-.05em{\sc i\kern-.025em b}\kern-.08em
T\kern-.1667em\lower.7ex\hbox{E}\kern-.125emX}}

\newcommand{\Ks}{\tilde{K}}
\newcommand{\Uub}{\Uc^*}
\newcommand{\Ulb}{{\Uc}_{\text{d}}}

\begin{document}

\vspace{-3cm}
\title{Outage Efficient Strategies for Network MIMO with Partial CSIT}

\author{Mari Kobayashi$^1$, \qquad
Sheng Yang$^1$, \qquad
M\'erouane Debbah$^1$, \qquad
Jean-Claude Belfiore$^2$
}

\maketitle
\noindent
{\small $^1$ SUPELEC, Gif-sur-Yvette,  91192, France \\
$^2$ T\'el\'ecom ParisTech, 75013, Paris, France}
\begin{abstract}
We consider a multi-cell MIMO downlink (network MIMO) where $B$
base-stations (BS) with $M$ antennas connected to a central station
(CS) serve $K$ single-antenna user terminals (UT).
Although many works have shown the potential benefits of network MIMO,
the conclusion critically depends on the underlying assumptions such as channel state information at transmitters (CSIT) and backhaul links.
In this paper, by focusing on the impact of partial CSIT, we propose an outage-efficient strategy.
Namely, with side information of all UT's messages and local CSIT, each
BS applies zero-forcing (ZF) beamforming in a distributed manner. For a
small number of UTs~($K\leq M$), the ZF beamforming creates $K$ parallel
MISO channels. Based on the statistical knowledge of these parallel
channels, the CS performs a robust power allocation that simultaneously
minimizes the outage probability of all UTs and achieves a diversity
gain of $B(M-K+1)$ per UT. With a large number of UTs~($K \geq M$), we
propose a so-called distributed diversity scheduling~(DDS) scheme to select a subset of $\Ks$ UTs with limited backhaul communication. It is proved that DDS achieves a diversity gain of $B\frac{K}{\Ks}(M-\Ks+1)$, which scales optimally with the number of cooperative BSs $B$ as well as UTs. Numerical results confirm that even under realistic assumptions such as partial CSIT and limited backhaul communications, network MIMO can offer high data rates with a sufficient reliability to individual UTs.
\end{abstract}

\newpage
\section{Introduction}
\begin{figure}[t]
	\begin{center}
	\epsfxsize=3in
	\epsffile{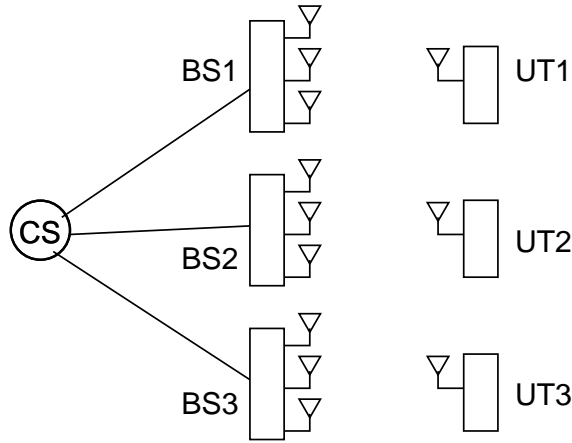}
	\end{center}
	\caption{A multicell downlink with $B=3$ BSs and $K=3$ UTs.}
	\label{fig:Model}
\end{figure}

Recently, network MIMO schemes, where neighboring Base-Stations (BSs) are
connected to form an antenna array, have been proposed as a means to drastically increase the downlink capacity and solve the interference management problem of cellular systems \cite{karakayali2006network}.
Inspired by this result, we consider the multi-cell MIMO downlink where $B$ BSs with $M$ antennas connected to a Central Station (CS) communicate simultaneously with $K$ User Terminals (UTs) with a single antenna each. Fig.\ref{fig:Model} illustrates an example of a multi-cell downlink system for $B=K=3$ and $M=3$. The channel at hand is modeled by the Multi-Input Single-Output (MISO) interference channel, defined by
\begin{eqnarray}\label{model}
	y_k[t] &=& \sum_{i=1}^B\hv_{ik}^H \xv_i[t] + n_k[t]
\end{eqnarray}
for $t=1,\dots, T$, where $y_k[t]$ is the channel output at UT $k$, $\hv_{ik}^H$ denotes the channel vector from BS $i$ to UT $k$,
$n_k[t]\sim\Nc_{\Cc}(0,1)$ is the Additive White Gaussian Noise (AWGN), and
$\xv_i\in\CC^{M\times 1}$ denotes the input vector transmitted by BS $i$ subject to the average power constraint $P_i$. 

If the CS or equivalently all the BSs have perfect \emph{Channel State Information at the Transmitter} (CSIT) and share the messages of all UTs, the channel at hand falls down into a classical $BM\times K$ MIMO broadcast channel with per-BS power constraints. In this case,
the optimal strategy to maximize the multi-cell throughput is joint dirty-paper coding \cite{karakayali2006network,shamai:ecd,somekh2007src}. In order to capture the essential features of the multi-cell systems while enabling the analysis tractable, the Wyner model \cite{wyner1994sta} has been widely considered in the literature. In \cite{somekh-information}, the authors provide a survey on the information theoretic results on the multi-cell systems under the Wyner model for the Gaussian and fading channels with a single-antenna BS ($M=1$). These include the per-cell downlink capacity based on the circular Wyner model \cite{somekh2007src,jing2007dmd,somekh2006dmc} and the corresponding analysis in the different asymptotic regime such as high SNR, a large number of BSs, UTs \cite{somekh2007src,jing2007dmd}.
If each BS is equipped with multiple antennas ($M>1$), the multi-cell downlink capacity can be naturally enhanced by exploiting the spatial degrees of freedom (see for example \cite{MMK,caire2008allerton,choi2006cgb}). 
For a small $B$, the MIMO multi-cell downlink channel is also referred
to the MIMO interference channel or MIMO-X channel under various message
sharing assumptions (see \cite{MMK,JafarShamai} and references therein). In these contributions, the sum degrees of freedom has been extensively studied.

Unfortunately, the global joint processing at the CS is difficult (if not impossible) in practice.
This calls for practical network MIMO schemes which build on distributed processing at each BS by explicitly taking into account realistic aspects.
A large number of recent contributions have been focused on practical designs by considering the following main limitations in network MIMO (see e.g. \cite{JakobISIT2010,simeone2009downlink,shamai2007cmc,marsch2007fod,Tamaki2007,Ng_Evans,Levy2009} and references therein).
First, a substantial amount of resource needs to be dedicated for the CS to obtain accurate CSI. In particular, this overhead increases significantly with the number of cooperative BSs, which in turn leaves few resource for the data transmission within a fixed coherence block.
In \cite{JakobISIT2010}, such tradeoff between the benefits of network MIMO and the overhead in channel estimation
has been studied for the case of the uplink with $M=1$ and $B=K$.
Second, the backbone links between the BSs and the CS are typically imperfect. They might be the capacity-limited \cite{simeone2009downlink,shamai2007cmc,marsch2007fod}, erroneous, or delayed \cite{Tamaki2007}.
This backhaul imperfectness will prevent the BSs from fully sharing the side
information on the messages or CSI. Similar effects may occur when only
adjacent BSs are connected to each other and exchange side information \cite{Ng_Evans,Levy2009}.

Our contribution is no exception. We aim to design a practical scheme which ensures high data rates with sufficient reliability to individual UTs under partial CSIT. The objective at hand is relevant to most of the current/next wireless standards \cite{sesia2009lte,WiMAX07}. To this end, we assume that each BS $i$ has local CSIT while the CS has only statistical CSIT. Notice that the last assumption is reasonable because 
the CS needs to track the downlink channels at a rate much slower than the coherence time.
We consider that the CS generates the messages destined to all $K$ UTs and sends them to the $B$ BS over the backbone link so that each BS locally encodes the messages and transmits. In order to concentrate on the impact of partial CSIT, we do not take into account other practical limitations. In particular, the underlying backbone
links are assumed to be perfect such that $B$ BSs can fully share the messages\footnote{If some backbone links are in failure, the corresponding BSs fail to encode the messages to $K$ users. This will reduce the effective number of BSs.}.
First, we address the case where the number of UTs is smaller than the
number of transmit antennas, i.e., $K\leq M$. In this case, each BS applies ZF beamforming
in a distributed manner, which creates $K$ parallel MISO channels. Based
on the statistical knowledge of these parallel channels, the CS performs
a robust power allocation that simultaneously minimizes the outage probability
of $K$ UTs and achieves a diversity gain of $B(M-K+1)$ per UT. Next, we consider a more relevant case of a
large number of UTs~($K > M$). In this case, we propose a simple
scheduling scheme called distributed diversity scheduling~(DDS) which efficiently
chooses a set of $\Ks \leq M$ UTs while limiting the amount of the backhaul communication.
More precisely, each BS $i$ with local CSIT knowledge
chooses its best set of UTs over the predefined partition and reports the corresponding index and value to the CS.
The CS then decides and informs the selected set to all BSs. Finally, the selected UTs are served exactly in the same manner as the previous case of $K\leq M$. It is proved that DDS offers a diversity gain of
$B\frac{K}{\Ks}(M-\Ks+1)$ to each UT and moreover this gain scales optimally in $B, K$, and $M$, respectively.
We remark that a similar CSIT assumption (local CSIT at BSs and statistical CSIT at the CS) has been also considered in \cite{Bjornson_Zakhour}. However, the objective of this contribution is maximizing the multi-cell throughput rather than minimizing the outage probability as we address here.

Numerical results validate the analysis in terms of diversity gain and show that our proposed distributed ZF beamforming significantly outperforms the non-cooperative scheme.  Namely, the outage performance of our proposed scheme improves with the number of cooperative BS, transmit antennas, and UTs. It is also shown in a simple one-dimensional topology that the performance gain can be emphasized
as the path-loss between neighboring cells decreases. The main finding of this paper is that
even in a realistic scenario with partial CSIT network MIMO can be beneficial by providing high data rate with a sufficient reliability to individual UTs. Such merit of network MIMO has been somehow overlooked in most of existing works assuming perfect CSIT.

The rest of the paper is organized as follows. The following section describes the system model. In Section \ref{sect:powerallocation} we present the power allocation policies that minimize the outage probability both under perfect and statistical CSI at the CS. We consider the relevant
case of $K\geq M$ in Section~\ref{sec:scheduling} where we propose a simple user scheduling algorithm and provide its diversity analysis.
Section \ref{sect:Results} shows some numerical results and Section
\ref{sect:conclusions} concludes the paper.

\textbf{Notations: } Throughout the paper, we use boldface lower case
letters to denote vectors, boldface capital letters to denote matrices.
$(\cdot)^\star$, $(\cdot)^T$, and $(\cdot)^H$ respectively denote the
complex conjugation, matrix transposition, and Hermitian transposition
operations.  $\Id_n$ and $\zerov_{n\times m}$ represent the $n\times n$
identity matrix and $n\times m$ zero matrix. The determinant, rank, trace, and Frobenius norm of a matrix $\Am$ are denoted by $|\Am|$, $\rank(\Am)$, $\trace(\Am)$, and $\|\Am\|^2_{\text{F}}$, respectively.
The dot-equality stands for the near-zero equality, that is, $f(\epsilon)\doteq \epsilon^n$ means $\lim_{\epsilon\to 0}\frac{\log f(\epsilon)}{\log \varepsilon} = n$. We let $\chi^2_{m}$ denote the chi-square distribution with $m$ degrees of freedom.

%
\section{Distributed Zero-Forcing Beamforming}

In order to split the processing at the CS and the BSs, it is reasonable to assume that the CS generates the messages and performs the resource allocation whereas each BS encodes and then transmits the symbols in a distributed fashion. More precisely,
we assume that the CS broadcasts $K$ messages intended to $K$ UTs which enable the BSs to cooperate in transmission. Then, each BS $i$ encodes these messages into $KT$ symbols $\{s_{ik}[t]\}$ for $k=1,\dots,K$ and $t=1,\dots, T$ by some capacity-achieving space-time coding (with a sufficiently large $T$ channel uses).  Under this setting, this section focuses on a distributed precoder design at each BS which requires only local CSIT $\{\hv_{ik}\}_{k=1}^K$ for any $i$.

In the multi-cell downlink channel (\ref{model}), we model the vector $\hv_{ik}$ of channel coefficients from BS $i$ to UT $k$ Gaussian distributed $\sim\Nc_{\Cc}(\zerov,\sigma_{ik}\Id_M)$ where the variance $\sigma_{ik}$ captures the path-loss of the corresponding link assuming that the UTs are arbitrary distributed. Furthermore, $\{\hv_{ik}\}$ are assumed to be i.i.d. over any pair $i,k$. We start with the definition of zero-forcing beamforming vectors as well as a useful lemma on the resulting channel statistics.
\begin{definition}[Zero-forcing beamforming vectors]
For a channel matrix $\Hm\in\CC^{K \times M}$ with $K \leq M$ linearly independent row vectors $\hv_1^H, \dots,\hv_K^H$,
there exists a zero-forcing beamforming matrix $\Gm=[\gv_1,\dots,\gv_K]\in\CC^{M \times K}$, composed by $K$ column vectors $\gv_1,\dots,\gv_K$, which is defined as
  \begin{equation}
    \Gm = \Hm^+ \diag\left\{ {a_k} \right\} \label{eq:def_BF}
  \end{equation}%
  where $\Hm^+ = \Hm^H \left( \Hm \Hm^H \right)^{-1}$ and
  $\diag\left\{ {a_k} \right\}$ is a diagonal matrix that normalizes the
  norm of the columns of $\Hm^+$ such that $\|\gv_k\|=1$ for any $k$.
\end{definition}%
\begin{lemma}\label{lemma:chi2}
If $\Hm \in \CC^{K \times M}$ has i.i.d. entries where each row $\hv_k^H\sim\Nc_{\Cc}(\zerov,\sigma_k \Id )$, then
$|a_k|^2$ as defined in (\ref{eq:def_BF}) is $\chi^2_{2(M-K+1)}$ distributed with mean $\EE[|a_k|^2]=\sigma_k$, $\forall\,k$.
\end{lemma}

We consider a simple ZF beamforming which enables each UT to achieve a multiplexing gain of one. At each channel use $t$, BS $i$ applies ZF beamforming to transmit $K$ symbols in a distributed manner. Due to the symmetry over all channel uses, we ignore the index $t$ hereafter. We form  the transmit vector of BS $i$ at any channel use as
\begin{equation}\label{x_i}
\xv_i = \Gm_i \uv_i =\sum_{k=1}^K \sqrt{p_{ik}}\gv_{ik} s_{ik} 
\end{equation}
where we let $\Gm_i=[\gv_{i1}, \dots,\gv_{iK}]$ be the ZF beamforming matrix corresponding to the channel matrix from BS
$i$ formed by row vectors $\hv_{i1}^H, \ldots, \hv_{iK}^H$, $\uv_i=[\sqrt{p_{i1}}s_{i1},\dots,\sqrt{p_{iK}}s_{iK}]^T$ is the vector of symbols where $s_{ik}\sim\Nc_{\Cc}(0,1)$ denotes the symbol
transmitted by BS $i$ to UT $k$ with power $p_{ik}$.
With this beamforming, the received signal at UT $k$ is given by
\begin{eqnarray}\label{MISO}
y_k= \sum_{i=1}^B \sqrt{p_{ik}} a_{ik} s_{ik} + n_k
\end{eqnarray}
where $a_{ik}=\hv_{ik}^H\gv_{ik}$ denotes the overall channel from BS
$i$ to UT $k$ and coincides with the definition in (\ref{eq:def_BF}) for each $i$.
From Lemma~\ref{lemma:chi2} and the independence of $\{a_{ik}\}$ over $i$, we remark that the original $B\times K$ MISO interference
channel (\ref{model}) is decoupled into $K$ parallel
$B\times 1$ MISO channels.

\begin{lemma} \label{lemma:diversity}
  Let $\yv = \Hm \xv + \zv$ denote a MIMO slow fading channel with
  standard complex Gaussian noise $\zv$. If we have
  \begin{align}
    \Pr\left\{ \| \Hm \|^2_{\text{F}} < \epsilon \right\} \doteq
    \epsilon^{d} \label{eqn:near-zero}
  \end{align}%
then the maximum diversity order of the channel is $d$.
\end{lemma}

\begin{theorem}\label{thm:miso}
  With distributed ZF beamforming, the diversity order of each UT is $B(M-K+1)$.
\end{theorem}

\begin{proof}
Appendix \ref{proof:MISO}.
\end{proof}

Assuming that each UT $k$ perfectly knows the channel state $\av^k = (a_{1k},\dots,a_{Bk})$, it decodes the space-time code and achieves the following rate
\begin{eqnarray}\label{MISOrate}
R_k = \log\left(1 + \sum_{i=1}^B|a_{ik}|^2 p_{ik} \right).
\end{eqnarray}
The capacity region of the $K$ parallel MISO channels (\ref{MISO}) for a \emph{fixed} set of power $\pv^k =(p_{1k},\dots,p_{Bk})$ and channel state $\av^k=(a_{1k},\dots,a_{Bk})$ for all $k$ is given by
\begin{eqnarray}
\Rc(\av;\pv) =
\left\{\Rm\in \RR_+^K : R_k  \leq \log\left(1 + \sum_{i=1}^B|a_{ik}|^2 p_{ik} \right)\;\forall k \right\}
\end{eqnarray}
where we let $\av=\{\av^k\}, \pv=\{\pv^k\}$ for a notation simplicity. The above region is clearly convex (rectangular for $K=2$). The capacity region of the $K$ parallel
MISO channels (\ref{MISO}) under the individual BS power constraints $\Pm=(P_1,\dots,P_B)$ for a \emph{fixed} channel state $\av$ is given by
\begin{eqnarray}
\Cc(\av;\Pm) = \bigcup_{\Pc\in\Fc} \Rc(\av;\pv)
\end{eqnarray}
where $\Pc$ denotes a power allocation policy $\av\longmapsto\pv$ that maps the channel state $\av$ into the power vector $\pv$ with component $\Pc_{ik}(\av)=p_{ik}$, $\Fc$ denotes the feasibility set satisfying $\sum_{k=1}^K \Pc_{ik}(\av) \leq P_i, \forall i$ for any channel realization $\av$.
The capacity region $\Cc(\av;\Pm)$ is convex and its boundary can be explicitly characterized by solving the weighted sum rate maximization as specified in subsection \ref{subs:PerfectCSI}.

\section{Power allocation minimizing outage probability}\label{sect:powerallocation}

In the previous section, we considered the distributed processing at each BS assuming that the power allocation is already done at the CS and each BS is informed about the resulting power partition. Here, we address the power allocation problem solved at the CS. In particular, we are interested in a robust power allocation strategy which requires only statistical CSIT. To this end, we consider a relevant scenario where the system imposes a set of target rate to each UT depending on its application. This is of typical of the current standards such as WiMax \cite{WiMAX07} and LTE \cite{sesia2009lte}. One of the important goals in such a situation is to minimize the outage probability simultaneously for all UTs. In order to formalize the problem, we let $\gamma_k$ denote the target rate of UT $k$ and form the target rate tuple   $\gammav=(\gamma_1,\dots,\gamma_K)$.
For a given $\gammav$, we define the outage probability as the average probability that $\gammav$ is not fulfilled by all $K$ UTs simultaneously, namely
\begin{eqnarray} \nonumber
P_{\rm out}(\gammav)&\eqdef & 1 - \EE_{\av}[\onev\{ \gammav  \in \Cc(\av,\Pm)\}]  \\ \label{Outage}
&=& 1- \Pr\left(   \gammav  \in \Cc(\av,\Pm) \right).
\end{eqnarray}
This section provides the power allocation policies minimizing the outage probability defined above under perfect and statistical CSIT.

\subsection{Perfect CSI at CS}\label{subs:PerfectCSI}
We start with a special case where the CS has perfect CSIT. In this case, we are particularly interested in the power allocation policy that provides the rate tuple proportional to the target rate tuple (rate-balancing). As seen shortly, this policy equalizes the individual outage probability of all UTs and thus provides the strict fairness among UTs. The latter is one of the most desired properties. Our objective is find the set of $\{p_{ik}\}$ satisfying
\begin{equation}\label{Prop}
 \frac{R_k(\pv)}{R_1(\pv)} = \frac{\gamma_k}{\gamma_1}\eqdef \alpha_k, \;\; k=2,\dots,K
\end{equation}
where we defined $\frac{\gamma_k}{\gamma_1}=\alpha_k$ and $\alpha_1=1$.
More precisely, the power allocation is a solution of \cite{lee:scm}
\begin{equation}\label{MinMax}
	\min_{\sum_k \theta_k=1} \max_{\Rm\in \Cc(\av,\Pm)} \sum_{k=1}^K
        \theta_k \frac{R_k}{\alpha_k}.
\end{equation}
Notice that the inner problem is the weighted sum rate maximization for a fixed weight set, while the outer problem with respect to $\theta_2,\dots,\theta_K$ is a convex problem for which a $(K-1)$-dimensional subgradient method can be suitably applied \cite{Bertsekas}. First, we describe a numerical method to solve the inner problem, given by
\begin{eqnarray}\label{WSRM}
\max_{\sum_{k=1}^Kp_{ik}\leq P_i,\forall i}  \sum_{k=1}^K w_k \log\left(1 + \sum_{i=1}^B|a_{ik}|^2 p_{ik} \right)
\end{eqnarray}
where $w_k=\theta_k/\alpha_k$ denotes a non-negative weight for all $k$ that we assume given for the time being. Since the above problem is convex (the objective function is concave and the constraints are linear), it is necessary and sufficient to solve the KKT condition \cite{ConvexBook}, given by
\begin{eqnarray}\label{KKT}
	 \frac{w_k |a_{ik}|^2}{1 + \sum_{j=1}^B|a_{jk}|^2
         p_{jk}}=\frac{1}{\mu_i}, \quad k=1,\dots,K,\ i=1,\dots,B
\end{eqnarray}
where $\mu_i$ denotes the Lagrangian variable to be determined such that $\sum_{k} p_{ik}\leq P_i$.
Unfortunately solving (\ref{KKT}) directly seems intractable. Nevertheless, the following waterfilling approach inspired by the iterative multiuser waterfilling for the MIMO multiple access channel \cite{YuWF} solves the KKT conditions iteratively.

{\bf Algorithm A1 : Iterative waterfilling algorithm for $B$-BS weighted sum rate maximization }
\begin{enumerate}
	\item Initialize $\pv_i^{(0)}=\zerov$ for $i=1,\dots,B$ and $c_{ik}^{(0)}=0$ for all $i,k$.
	\item At each iteration $n$ \\
	For $i=1,\dots,B$
        \begin{itemize}
          \item Compute $c_{ik}^{(n)} = \sum_{j\neq i}|a_{jk}|^2 p_{jk}^{(n)}$
          \item Waterfilling step : let $\pv_i^{(n)}$ be
          \begin{eqnarray}\label{WF}
    p_{ik}^{(n)} &=& \left[w_k \mu_i -\frac{1+ c_{ik}^{(n)}}{|a_{ik}|^2}\right]_+, \;\; \forall k
\end{eqnarray}
where $\mu_i$ is determined such that $\sum_{k=1}^Kp_{ik}^{(n)}=P_i$.
        \end{itemize}
	End
	\item Continue until convergence
\end{enumerate}

We have the following remarks on the proposed algorithm.
\begin{remark}
Algorithm A1 is a generalization of the classical waterfilling algorithm for the $K$ parallel channels under the total power constraint to the case with $B$ transmitters with individual power constraints.  Indeed, for a single-BS case ($B=1$), the objective function reduces to the weighted sum rate of the $K$ parallel non-interfered channels given by
\begin{eqnarray}\label{B1}
    \max_{\qv\geq \zerov, \sum_k q_{k}\leq P}\sum_{k=1}^K
    w_k\log\left(1+ |a_{k}|^2 q_{k}\right).
\end{eqnarray}
\end{remark}

\begin{remark}
The set of powers in (\ref{WF}) correspond to the solution to the new objective function
\begin{eqnarray}\label{NewObj}
 \pv_i^{(n)}= \arg \max_{\qv\geq \zerov, \sum_k q_{k}\leq
 P_i}\sum_{k=1}^K w_k\log\left(1+ \frac{|a_{ik}|^2
 q_{k}}{1+c_{ik}^{(n)}}\right).
\end{eqnarray}
It can be easily seen that this new problem and the original problem
(\ref{WSRM}) yield the same KKT conditions (\ref{KKT}) for
$c_{ik}^{(n)}=\sum_{j\neq i}|a_{jk}|^2 p_{jk}^{(n)}$. In other words,
the solution (\ref{WF}) of BS $i$ corresponds to treating
$\{c_{ik}\}_{k=1}^K$, function of the powers $\{\pv_{j}\}_{j\neq i}$ of
the other BSs, as additional noise (constant), i.e., as if they did not
depend on $\pv_i$. Under the individual power constraints for each BS, the sequential iteration over different BSs probably converges (see the convergence proof below).
\end{remark}

By a rather straightforward extension of \cite[Theorem 1, Theorem 2]{YuWF}, we have the following convergence result.
\begin{theorem}\label{theorem:waterfilling}
 Algorithm A1 converges to the optimal solution of the weighted sum rate maximization (\ref{WSRM}).
\end{theorem}

\begin{proof}
The proof follows in the footsteps of the proofs of \cite[Theorem 1, Theorem 2]{YuWF}.
At each iteration, the proposed iterative algorithm finds the solution to the weighted sum rate maximization of the single-BS parallel channels (\ref{B1}) for each BS while treating the powers of all the other BSs as additional noise. Comparing the objective of the single-BS parallel channels (\ref{B1}) and that of the multi-BS parallel channels (\ref{NewObj}), they differ only from the constant $c_{ik}(\{p_{jk}\}_{j\neq i})$. Hence, the objective is nondecreasing at each waterfilling step and further converges to a limit. The powers $\pv_1,\dots,\pv_B$ of $B$ BSs also converge to the limit $\pv^{\star}_1,\dots,\pv^{\star}_B$. This is because the solution to the single-BS parallel channels is unique.
\end{proof}

The outer problem consists of minimizing the solution of the inner problem with respect to $\theta_2,\dots,\theta_K$. This can be done by the subgradient method. We summarize the overall algorithm as follows.\\
{\bf Rate balancing algorithm }
\begin{enumerate}
	\item Initialize $\theta_k^{(0)}\in [0,1]$ for $k>1$.
	\item At iteration $n$, compute
	\begin{equation}
		\Rm^{(n)} =\arg\max_{\Rm\in \Cc(\av,\Pm)}\sum_{k=1}^2 \theta_k^{(n)} \frac{R_k}{\alpha_k}
	\end{equation}	
	via waterfilling approach.\\
    Compute a subgradient of user $k$
	\[ \Delta_k^{(n)}=\frac{R_k^{(n)}}{\alpha_k}-R_1^{(n)}. \]
	\item Update the weights by subgradient :
\[\thetav^{(n+1)}=\thetav^{(n)}- s_n \Deltam^{(n)} \]
where $s_n$ denotes the step size at iteration $n$ that can be chosen according to a decreasing rule.
\end{enumerate}

The overall algorithm, the inner problem solved by algorithm A1 and the outer problem solved by the subgradient method, implements the rate-balancing by allocating the rates proportional to the target rate tuple. Hereafter, we let
$R_k^{\star}(\av)$ denote the rate of user $k$ allocated by the rate-balancing algorithm in channel state $\av$. Notice that the achievable rate (\ref{MISOrate}) is a deterministic function of $\Pc^{\star}(\av)$. Then, we have the following result.

\begin{theorem}\label{theorem:RateBalancing}
The rate-balancing power allocation minimizes the outage probability. Moreover, it equalizes the individual outage probability of all $K$ UT by letting $\Pr(\gamma_1<R_1^{\star}(\av))=\dots=\Pr(\gamma_k<R_k^{\star}(\av))$.
\end{theorem}

\begin{proof}
Appendix \ref{proof:RateBalancing}.
\end{proof}

\vspace{-0.9em}
\subsection{Statistical CSI at CS}
We consider a more realistic case when the CS has only statistical knowledge of the equivalent channels $\av$. Formally, we define
the power allocation policy $\Pc^{\rm stat} : \sigmav \mapsto \pv$ as a function mapping the variances of channel coefficients $\sigmav =(\sigma_{11},\dots,\sigma_{BK})$ into the power vector $\pv$ with component $\Pc^{\rm stat}_{ik}(\sigmav)=p_{ik}$.
Since the equivalent channel coefficients $\{a_{ik}\}$ after ZF beamforming are correlated over different $k$ for each BS $i$, the individual outage events are dependent. Nevertheless, this dependency can be made arbitrarily small by a simple interleaving over frequency bands for example.
 Under this assumption, we approximate
the outage probability for a fixed power allocation $\pv$ as
\begin{eqnarray} \label{generalOutage}
    P_{\rm app-out}(\gammav,\pv)
    &= & 1- \prod_{k=1}^K\Pr\left( \Delta_k(\pv^k) > 2^{\gamma_k}-1  \right)
\end{eqnarray}
where we let $\Delta_k(\pv^k)=\sum_{i=1}^B|a_{ik}|^2 p_{ik}$.
First, we remark that for a fixed power $\pv^k=(p_{1k},\dots,p_{Bk})$ of UT $k$ , $\Delta_k(\pv^k)$ is a Hermitian quadratic form of complex Gaussian random variables given by
\begin{footnotesize}
\begin{eqnarray*}
\Delta_k(\pv^k) &=& (a_{1k}^* ,\dots, a_{Bk}^*)
 \left(\begin{array}{ccc}
 p_{1k} &  & \zerov \\
 & \ddots & \\
 \zerov &  & p_{Bk}  \\
 \end{array}\right)
 \left(\begin{array}{c}
 a_{1k}\\\vdots \\ a_{Bk} \end{array}\right) \\ \nonumber
&= &(\wv_{1k}^H \dots \wv_{Bk}^H) \left(\begin{array}{ccc}
 p_{1k}\Id & & \zerov \\
 & \ddots & \\
 \zerov & & p_{Bk}\Id\\
 \end{array}\right)
 \left(\begin{array}{c}
 \wv_{1k} \\ \vdots \\ \wv_{Bk} \end{array}\right) 
\end{eqnarray*}
\end{footnotesize}
where the second equality follows by replacing a chi-square random variable $|a_{ik}|^2\sim\chi^2_{2(M-K+1)}$ with $||\wv_{ik}||^2$
where $\wv_{ik}\sim\Nc_{\Cc}(\zerov, \frac{\sigma_{ik}}{M-K+1}\Id_{M-K+1})$.
The individual outage probability that UT $k$ cannot support $\gamma_k$ for a fixed $\pv^k$ is
\begin{equation}\label{cdf}
    \Pr(\Delta_k(\pv^k)\leq c_k) = \frac{1}{2\pi j}\int_{c-j\infty}^{c+j\infty} \frac{e^{s c_k}\Phi_{\Delta_{k}(\pv^k)}(s)}{s}ds
\end{equation}
where we let $c_k= 2^{\gamma_k}-1$ denote the target SNR of UT $k$, 
and the Laplace transform of $\Delta_k(\pv^k)$ is given by
\begin{eqnarray}\label{laplace}
    \Phi_{\Delta_k(\pv^k)} (s) 
    = \prod_{i=1}^B\frac{1}{(1+s \frac{p_{ik}\sigma_{ik}}{M-K+1})^{M-K+1}}.
\end{eqnarray}
From (\ref{laplace}), we see immediately that each UT achieves a diversity gain of $B(M-K+1)$, which agrees well with Theorem \ref{thm:miso}.
The widely used upper bound is the Chernoff bound for fixed powers $\pv^k$ given by
\begin{equation}\label{Chernoff}
   \Pr(\Delta_k(\pv^k)\leq c_k)\leq \min_{\lambda\geq 0} e^{\lambda c_k}\Phi_{\Delta_{k}(\pv^k)}(\lambda) \eqdef \overline{F}(c_k,\pv^k). 
\end{equation}
Using the last expression of the Chernoff upper bound for each $k$, the approximated outage probability for a fixed $\pv$ is upper-bounded by
\begin{eqnarray} \label{JointChernoff}
    P_{\rm app-out}(\gammav,\pv) 
    &\leq & 1- \prod_{k=1}^K (1-\overline{F}(c_k,\pv^k)) 
\end{eqnarray}
Since the power optimization based on the exact outage probability is not amenable, we search the power allocation that minimizes the Chernoff upper bound, equivalently solves
\begin{eqnarray}\label{objB}
    \mbox{maximize} &&  f(\{\lambda_k\},\{p_{ik}\})\eqdef \prod_{j=1}^K(1-h_k(\lambda_k,\pv^k)) \\  \nonumber \nonumber
    \mbox{subject to } && \sum_{k=1}^K p_{ik} \leq P_i, \;\;\; i=1,\dots,B \\ \nonumber
    && \lambda_k \geq 0, \;\; k=1,\dots,K \\ \nonumber
    && p_{ik} \geq 0, \;\;\; i=1,\dots,B, k=1,\dots,K
\end{eqnarray}
where we define
\begin{eqnarray}
    h_k(\lambda_k,\pv^k) &=&\frac{e^{\lambda_k c_k}}{\prod_{i=1}^B (1+\beta_{ik}\lambda_k p_{ik} )^{M-K+1}}
\end{eqnarray}
where $\beta_{ik}=\frac{\sigma_{ik}}{M-K+1}$. If $h_k\geq 1$ for some
$k$, the objective becomes null regardless of the power allocation. In
this case a reasonable choice is to let $p_{ik}=0, \forall i$ for such
$k$ and equally allocate the power to $\{p_{ik'}\}$ for $k'\neq k$. In the following, by focusing on the case $h_k<1$ for all $k$, we provide an efficient numerical method to solve the problem (\ref{objB}). We remark first that the maximization of $f$ with respect to $\{\lambda_k\}$ can be decoupled into the minimization of $h_k$ over $\lambda_k$ for each $k$, where $h_k$ is convex in $\lambda_k$. Moreover, since $f$ is concave in $\{p_{ik}\}$, the overall problem is convex.

{\bf Minimization of $h_k$ over $\lambda_k$ }
It can be easily verified that $h_k$ is monotonically decreasing in $\lambda_k$. The optimal $\lambda_k$ for a fixed set of powers is the solution of
\begin{eqnarray}\label{lambdak}
    \frac{c_k}{M-K+1} =\sum_{i=1}^B \frac{\beta_{ik}p_{ik}}{1+\lambda_k \beta_{ik}p_{ik}}
\end{eqnarray}
which is a polynomial of degree $B$. For $B=2$, the solution is given in a closed form.
\begin{eqnarray}\label{lambda}
\lambda_k = \left\{
\begin{array}
[c]{ll}%
    \frac{M-K+1}{c_k}- \frac{1}{\beta_{1k}p_{1k}+\beta_{2k}p_{2k}}, & \quad\mbox{if $\sum_{j=1}^2\beta_{jk}p_{jk}=0$}\\
    \frac{-\left(\beta_{1k}p_{1k}+\beta_{2k}p_{2k}-\frac{2(M-1)\beta_{1k}p_{1k}\beta_{2k}p_{2k}}{c_k}\right)+
    \sqrt{(\beta_{1k}p_{1k}-\beta_{2k}p_{2k})^2+ \left(\frac{2(M-1)\beta_{1k}p_{1k}\beta_{2k}p_{2k}}{c_k}\right)^2}}{2\beta_{1k}p_{1k}\beta_{2k}p_{2k}}, & \quad\mbox{otherwise}
\end{array}
\right.
\end{eqnarray}

{\bf Maximization of $f$ over $p_{ik}$ }
Since $f$ is concave in $\{p_{ik}\}$, we form the Lagrangian function by introducing
$B$ Lagrangian multipliers $\{\mu_i\}$ each of which is associated to the power constraint of BS $i$.
By arranging the term common for all $k$, we obtain the KKT conditions for $k=1,\dots,K$
\begin{eqnarray}\label{KKTPowerB}
  && \frac{h_k(\pv^k)}{1-h_k(\pv^k)} \frac{\beta_{ik}\lambda_k}{1+
  p_{ik}\beta_{ik}\lambda_k}=\mu_i. 
\end{eqnarray}
When treating $p_{1k},\dots,p_{i-1,k},p_{i+1,k},\dots,p_{B,k}$ fixed, the LHS of (\ref{KKTPowerB}), denoted by
$\phi_{ik}$, is a strictly positive and monotonically decreasing
function of $p_{ik}$ (since we exclude the case $\lambda_k=0$). It
remains to determine $\mu_i$ such that the power constraint of BS $i$ is
satisfied , i.e., $p_{i1}+\dots+p_{iK}=P_i$. When treating the powers $\{\pv_j\}_{j\neq i}$ of the other BSs $j\neq i$ fixed, the powers $\pv_i$ of BS $i$ can be found by a simple line search of $\mu_i$.

The following summarizes our proposed iterative algorithm to minimize the Chernoff upper bound, equivalently solve (\ref{objB}).

\textbf{Algorithm A2 : iterative algorithm for the Chernoff upper bound minimization}
\begin{enumerate}
  \item Initialize $\pv^{(0)}$
  \item At iteration $n$\\
   For $i=1,\dots,B$
    \begin{itemize}
      \item Update $\lambdav^{(n)}$ by solving the polynomial (\ref{lambdak})
      \item Find the new power vector $\pv_{i}^{(n)}$ of BS $i$ by line search
    \end{itemize}
   End
  \item Continue until converge $f$
\end{enumerate}

Although we are unable to provide a formal proof, we conjecture that Algorithm A2 converges to its optimal solution. At each iteration, $\lambda_k^{(n)}$ is determined as a unique solution for all $k$ and a fixed set of powers. Regarding the power iteration, since the objective (\ref{KKTPowerB}) is concave in $p_{ik}$ when fixing all other powers, a sequential update of the powers
$\pv_1, \pv_2, \dots,\pv_B, \pv_1...$ shall converge under individual BS power constraints by a similar argument as the proof of Theorem \ref{theorem:waterfilling}.

\section{Effect of user scheduling}\label{sec:scheduling}

In this section we address the relevant case when the number of UTs is larger than the number of transmit antennas $K > M$. In order for each BS to apply the ZF beamforming in a distributed fashion, a set of $\Ks \leq M$ UTs shall be selected beforehand. We assume that the user selection (scheduling) is handled by the CS together with the power allocation. In particular, we focus on a user selection method which achieves a high diversity order while limiting the amount of the side information
necessary at the CS and the BSs. In the following, we present our proposed user selection scheme as well as the analysis on its achievable diversity gain.

\subsection{Distributed Diversity Scheduling (DDS)}
%

Let $\Sc, \Uc$ denote the set of all $K$ users, the $\Ks$ selected users, with $|\Sc| = K$, $|\Uc| = \Ks$, respectively. Let us also define
$\Qc(\Ks)$ as the set of all possible user selections, i.e., $\Qc(\Ks) = \left\{ \Uc\ |\
\Uc\subseteq\Sc,\ |\Uc| = \Ks \right\}$ for $\tilde{K}\leq M$.\footnote{For
convenience of notation, we will drop the argument $\Ks$ whenever
confusion is unlikely.}
Then, the equivalent channel from the BSs to the selected users is
\begin{equation}
  y_k = \av^k \uv^k + z_k, \quad
  k \in \Uc,
\end{equation}%
which is a MISO channel with $\av^k = \left[ a_{1k} \cdots a_{Bk}
\right]$ and $\uv^k = \left[ \sqrt{p_{1k}} s_{1k} \cdots \sqrt{p_{Bk}}
s_{Bk} \right]^T$. 
For convenience, we only consider the diversity order of the worst user
and refer it as the diversity of the system hereafter.
Since the diversity order of a given channel depends solely on the
Euclidean norm of the channel matrix, as shown in
lemma~\ref{lemma:diversity},  the following user selection scheme maximizes the diversity of the system
\begin{equation}
  \Uub = \arg_{\Uc}\max_{ \Uc \in \Qc } \min_{k\in\Uc} \|\av^{k}\|^2.
  \label{eq:Uub}
\end{equation}%
Unfortunately, this scheduling scheme has two major
drawbacks:~1)~perfect knowledge at the CS on $\{\av_k\}$, crucial for the
scheduling, is hardly
implementable as aforementioned, and 2)~the maximization over all $|\Qc(\Ks)|=\binom{K}{\Ks}$ possible sets
$\Uc$ grows in polynomial time with $K$.

To overcome the first drawback, we use the following selection scheme
\begin{equation}
  \Ulb = \arg_{\Uc}\max_{i=1\ldots B}
  \max_{{\Uc \in \Qc}} \min_{k\in\Uc}  |a_{ik}|^2.
\end{equation}%
That is, BS $i$ finds out the set $\Uc$ that maximizes
$\min_{k\in\Uc}  |a_{ik}|^2$ and sends both the index of the set and the
corresponding maximum value to the CS. Upon the reception of $B$ values
and the corresponding sets from the $B$ BS, the CS makes a decision by
selecting the largest one. Therefore, only partial
channel state information is communicated in the BS-CS link. To address
the second drawback, we narrow down the choices of $\Uc$ to the following $
\kappa = K/\Ks$ possibilities\footnote{Here, we assume that $K/\Ks$ is integer
for simplicity of demonstration. However, it will be shown that same
conclusion holds otherwise.}
$$ \Pc_\Sc = \left\{ \Uc_1, \Uc_2, \ldots, \Uc_\kappa \right\},\quad
\bigcup_i \Uc_i = \Sc,\ \Uc_i\cap\Uc_j=\emptyset,\ \forall\, i\neq j, \ |\Uc_i|=\Ks,\ \forall\,i. $$
In other words, $\Pc_\Sc$ is partition of the set of all users $\Sc$.
Furthermore, it is assumed that the partition $\Pc_\Sc$ is fixed
by the CS and known to all BSs.  Hence, the proposed scheduling scheme selects the following set of users
\begin{equation}
  \Ulb = \arg_{\Uc}\max_{i=1\ldots B}
  \max_{{\Uc \in \Pc_\Sc}} \min_{k\in\Uc}  |a_{ik}|^2.
  \label{eq:U_bar}
\end{equation}

To summarize, the scheduling scheme works as follows
\begin{enumerate}[  i)]
\item The CS fixes a partition $\Pc_\Sc$ and informs it to all BSs.
\item BS~$i$ finds $\max_{\Uc\in \Pc_\Sc} \min_{k\in\Uc}  |a_{ik}|^2$, and sends
  this value and the index of the maximizing set $\Uc$ to the CS.
\item The CS chooses the largest value and broadcasts the index of the
  winner set $\Ulb$ as defined in~(\ref{eq:U_bar}).
\item All the BSs serve simultaneously the UTs in $\Ulb$.
\end{enumerate}

\begin{figure}
    \begin{center}
   \epsfxsize=3.7in
   \epsffile{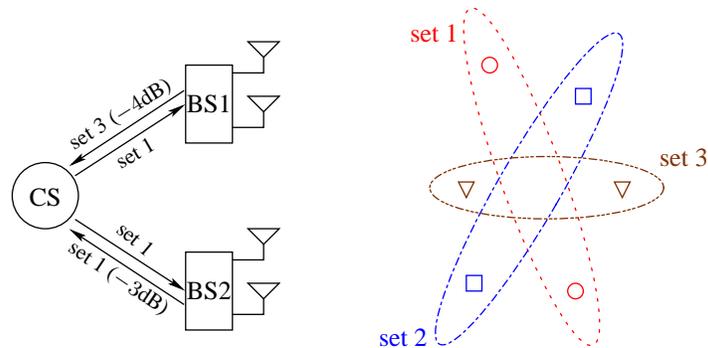}
    \end{center}
    \caption{An example scenario of user scheduling with two BSs and six
    UTs. }
    \label{fig:toy}
\end{figure}

An example of two BSs and six UTs is shown in Fig.~\ref{fig:toy}. In
this example, in order to serve two UTs simultaneously, a partition of
three sets is fixed by the CS. With local CSI, each BS compares the
coefficients $\min_{k\in\Uc}  |a_{ik}|^2$ for all three sets $\Uc$,
finds out the largest one, and sends the corresponding ``index(value)''
pair to the CS.  The CSI compares the values and broadcasts the index of
the winning set~(set $1$ in this example).

\subsection{Diversity gain analysis}
In this subsection, we analyze the diversity gain achieved by our proposed DDS and compare it with the upper bound. The result is summarized in the following Theorem.

\begin{theorem}\label{theorem:diversityDDS}
  Let $K$, $B$, and $M$ denote the number of UTs, number of BSs, and
  number of antennas per BS. By serving $\Ks$ UTs simultaneously, the
  following diversity gain is achievable with DDS
  \begin{equation}
    d_{\Ulb}(\Ks) \geq B \frac{K}{\Ks} \left( M- \Ks+1  \right) ,
    \label{eq:diversity_scheduling}
  \end{equation}%
  {for $K=n\Ks$ with an integer $n$}. Furthermore, the optimal diversity gain achieved by (\ref{eq:Uub}) is
  upper-bounded by
  \begin{equation}
    d_{\Uub}(\Ks) \leq B (K-\Ks+1) \left( M -\Ks+1  \right).
    \label{eq:diversity_scheduling_ub}
  \end{equation}%
\end{theorem}

\begin{proof}
Appendix \ref{proof:diversityDDS}.
\end{proof}

\begin{remark}
  For a fixed $\Ks$, the diversity gain of the proposed scheduling scheme
  grows as $O(BKM)$, i.e., the optimal diversity scaling with $B$, $K$,
  and $M$. In this sense, the proposed scheme is order optimal in terms
  of diversity gain. Also note that the lower bound
  (\ref{eq:diversity_scheduling}) and the upper bound
  (\ref{eq:diversity_scheduling_ub}) coincide in some specific settings.
  First, $d_{\Ulb}(1) = d_{\Uub}(1) = BKM$, $\forall\,B,K,M$. That is, the proposed scheme is
  diversity optimal if only one user is served in the system.
  Then, for $K\leq M$, we have $d_{\Ulb}(K) = d_{\Uub}(K) = B(M-K+1)$.
  This corresponds to the case where all users in the system are served
  simultaneously. 
\end{remark}

\begin{remark}
  Interestingly, exactly the same diversity order is achieved for
  ${\Uc}^*$ if we set $\Qc=\Pc_\Sc$.
  To see this, let us rewrite
  \begin{align}
    \min_{ k \in {\Uc}^*} \| \av^k \|^2
    &= \max_{\Uc\in\Pc_\Sc} \min_{k\in\Uc} \|\av^k\|^2 \\
    &\leq  \max_{\Uc\in\Pc_\Sc} \|\av^k\|^2 , \quad \forall\, k\in\Uc
  \end{align}%
  and that $\max_{\Uc\in\Pc_\Sc} \|\av^k\|^2 $ is of
  diversity ${B \frac{K}{\Ks} \left( M+1-\Ks \right)  }$.
\end{remark}
%

\begin{remark}
  When $\Ks$ does not divide $K$, we consider only $\left\lfloor\frac{K}{\Ks}\right \rfloor\Ks$
  out of $K$ users. Since $\Ks$ divides
  $\left\lfloor\frac{K}{\Ks}\right\rfloor \Ks$, the following diversity gain can be achieved
  \begin{equation}
    d_{\Ulb}(\Ks) = B \left\lfloor \frac{K}{\Ks} \right\rfloor \left( M-\Ks+1  \right)
    \label{eq:diversity_scheduling2}
  \end{equation}%
  with the proposed DDS.
\end{remark}

\section{Numerical Examples}\label{sect:Results}
This section provides some numerical examples to verify the behavior of our proposed distributed ZF beamforming scheme
in a simple network MIMO configuration with $B=2$ cooperative BSs. We assume the same power constraint at both BSs $P_1=P_2$ and let $P$
denote the SNR.

Fig. \ref{fig:AsymOutage} shows the outage probability performance versus SNR for $K=2$ and $M=2,4$. The target rate is fixed to $(\gamma_1,\gamma_2)=(3,1)$ bit/channel use, and we let $\sigma_{ik}=1$ for all $i, k$.
We compare the different power allocation strategies, algorithm A1 with perfect CSIT, algorithm A2 with statistical CSIT, and
equal power allocation ($p_{i1}=p_{i2}=P/2$ for $i=1,2$). For the sake of comparison, we also consider the case without
network MIMO (no message sharing) where each BS sends a message to its corresponding UT in a distributed fashion. In order to make the comparison fair in terms of complexity, we let each BS $i$ send the symbol $s_i$ by ZF beamforming, i.e. $\xv_i=\sqrt{P} \gv_i s_i$ where $\gv_i$ is a unit-norm vector orthogonal to $\hv_{ik}$ for $k\neq i$. From Lemma \ref{lemma:chi2},  such a system offers a diversity order of $M-K+1$ for each UT. As expected from Theorem \ref{thm:miso}, we observe that our BS cooperation schemes enables to achieve a diversity gain of $2(M-K+1)$, i.e. $2, 6$ with $M=2, 4$, respectively. These gains are twice as large as the case without network MIMO.
Moreover, the proposed algorithms provide a significant power gain compared to equal power allocation.

In Fig. \ref{fig:IndividualOutage}, we plot the individual outage probability such that each UT $k$ cannot support its target rate $\gamma_k$ under the same setting as Fig. \ref{fig:AsymOutage} for $M=2$.
With perfect CSIT, our proposed waterfilling allocation A1 guarantees the identical outage probability for both UTs by offering the strict fairness. This agrees well with the second part of Theorem \ref{theorem:RateBalancing}. Under statistical CSIT, algorithm A2 provides
a better outage probability to UT 1 but keeps the gap between two UTs smaller than the equal power allocation.

\begin{figure}
    \begin{center}
   \epsfxsize=5in
   \epsffile{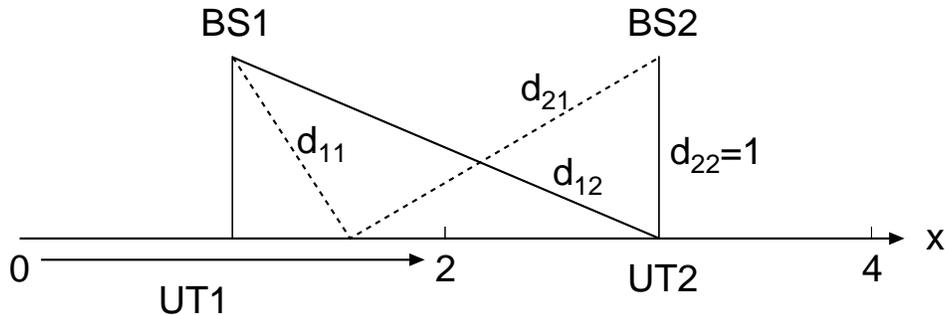}
    \end{center}
    \caption{One dimensional configuration.}
 \label{fig:Configuration}
\end{figure}

In order to evaluate the impact of asymmetric path loss on the outage performance, we consider a simple 1-D configuration illustrated in Fig. \ref{fig:Configuration} where UT 2 is located at $x=3$ and UT 1 moves from $x=0$ to $x=2$. Assuming that BS 1, 2 is at $x=1, 3$, respectively, we vary $d_{11}=\sqrt{1+(1-x)^2},d_{12}=\sqrt{1+(3-x)^2}$
while we fix the position of UT2 by letting $d_{12}=\sqrt{5},d_{22}=1$. By taking into account the path loss $\sigma_{ik}=d_{ik}^{-3}$, we plot the outage probability as a function of the position $x$ of UT 1 in Fig. \ref{fig:OutagevsDistance}. We consider $M=4$, SNR $P=10$ dB and fix the target rate $\gamma_1=\gamma_2=1$ bit/channel use.
We observe that the proposed distributed ZF scheme provides a significant gain compared to the case without network MIMO especially as UT 1 gets closed to the cell boundary ($x=2$). This is because the performance without network MIMO only depends on $d_{11}$ while the distributed ZF becomes more and more beneficial as $d_{21}$ decreases.

Finally, Fig. \ref{fig:OutageManyUsers} shows the outage probability
versus SNR when we have more users than the number of served users ,
i.e. $K\geq \Ks=2$. Considering the same setting as Fig.
\ref{fig:AsymOutage} for $M=4$, we apply distributed diversity scheme to
select a set of two users among $K=2,4,6$. Once the user selection is
done, any power allocation studied in Section \ref{sect:powerallocation}
can be applied. However, it is non-trivial (if not impossible) to characterize the statistics of the overall channel gains in the presence of any user scheduling. Hence, we illustrate here only the performance with equal power allocation under statistical CSIT. As a matter of fact, any smarter allocation shall perform between the waterfilling allocation and the equal power allocation. As expected from Theorem \ref{theorem:diversityDDS}, the diversity gain increases significantly as the number $K$ of users in the system gets large.

\section{Conclusions}\label{sect:conclusions}
We considered the multi-cell downlink system (network MIMO) where $B$
BSs, perfectly connected via the reliable backbone links to the CS, wish
to communicate simultaneously with $K$ UTs. As one of the realistic
limitations of network MIMO, we explicitly accounted for partial CSIT,
i.e. local channel knowledge at each BS and statistical channel
knowledge at the CS. Under this setting, we proposed an outage-efficient
strategy which builds on distributed ZF beamforming to be performed at
each BS and efficient power allocation algorithms at the CS. For the
case of a small number of users $K\leq M$, the proposed scheme enables
each UT to achieve a diversity gain of $B(M-K+1)$. For the case of many
users $K\geq M$, we proposed distributed diversity scheduling (DDS) which can be implemented in a distributed fashion at each BS and requires only limited amount of the backbone communications. We also proved that DDS can offer the diversity gain of $B\frac{K}{\Ks}(M-\Ks+1)$ and this gain scales optimally with the number of cooperative BSs as well as the number of UTs. The main finding is that limited BS cooperation can still make network MIMO attractive in the sense that a well designed scheme can offer high data rates with sufficient reliability to individual UTs. The proposed scheme can be suitably applied to any other interference networks where the transmitters can perfectly share the messages to all UTs and a master transmitter can handle the resource allocation.



\appendix
\subsection{Proof of Theorem~\ref{thm:miso}} \label{proof:MISO}

\begin{figure}[t]
	\begin{center}
	\epsfxsize=3.2in
	\epsffile{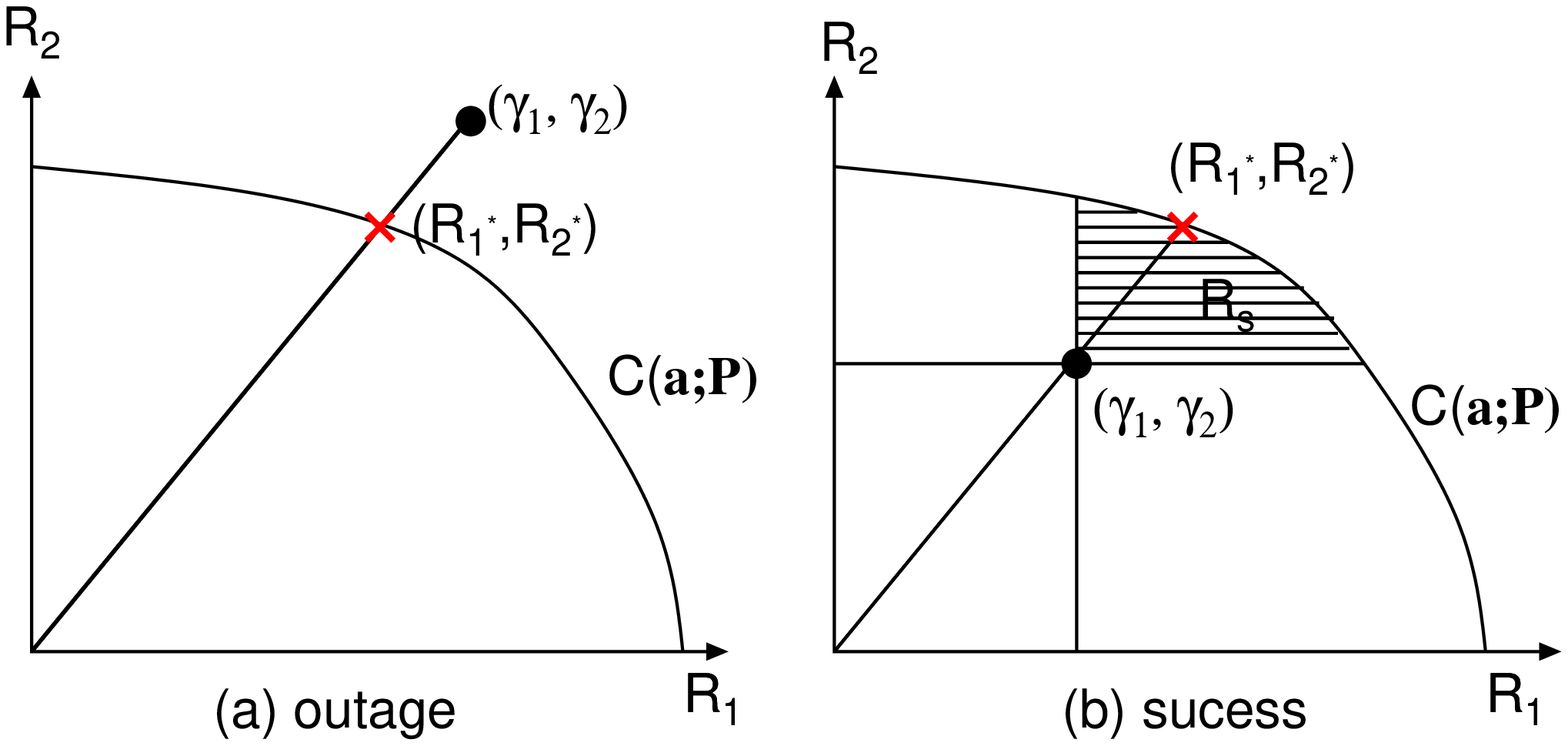}
	\end{center}
	\caption{Rate region $\Cc(\av;\Pm)$ and subregion $\Rc_s$}
	\label{fig:Optimality}
\end{figure}

  The channel (\ref{MISO}) is a MISO channel defined by $\av^k =
  (a_{1k},\dots,a_{Bk})$ and
  \begin{align}
    \Pr\left\{ \| \av^k \|^2 < \epsilon \right\}
    &\doteq \Pr\left\{ \max_i | a_{ik} |^2 < \epsilon \right\} \\
    &= \prod_i \Pr\left\{ | a_{ik} |^2 < \epsilon \right\} \\
    &= \left( \Pr\left\{ | a_{ik} |^2 < \epsilon \right\}
    \right)^B \\
    &\doteq \epsilon^{B(M-K+1)} \label{eqn:toto}
  \end{align}%
  From lemma~\ref{lemma:diversity}, the maximum diversity is
  $B(M-K+1)$. 

\subsection{Proof of Theorem~\ref{theorem:RateBalancing}}\label{proof:RateBalancing}
First we remark that for a given channel realization $\av$ the following two cases occur:
\begin{enumerate}[(a)]
\item The target rate tuple is outside the region $\gammav\notin\Cc(\av,\Pm)$
 \item The target rate tuple is inside the region $\gammav\in\Cc(\av,\Pm)$
\end{enumerate}
The above two cases are illustrated in Fig.\ref{fig:Optimality}(a), (b) respectively for $K=2$.
For the case (a), we are in an outage event regardless of the power allocation. For the case (b), we define the subregion $\Rc_s(\av)$
\begin{eqnarray}
   \Rc_s(\av) \eqdef \{\Rm | \Rm \in \Cc(\av,\Pm), R_k \geq \gamma_k, \;k=1,\dots,K \}
\end{eqnarray}
depicted in a shadow area in Fig. \ref{fig:Optimality} (b). We define $\Pc_s$ as a class of the power allocation policies that maps $\av$ into the rate tuple $\Rm$ inside $\Rc_s(\av)$ whenever we are in case (b).
We remark that any policy belonging to $\Pc_s$ results in a successful transmission, and thus 
minimizes the outage probability. Since the proposed rate balancing scheme allocates the power $(p_1^{\star},\dots,p_K^{\star})$ so that the rate-tuple $(R_1^{\star},\dots,R_K^{\star})$ is proportional to $\gammav$ on the boundary of $\Cc(\av, \Pm)$ whenever $\gammav\in\Cc(\av,\Pm)$, it belongs to the class $\Pc_s$. This establishes the first part.

We now prove the second part. It is immediate to see that with the rate balancing scheme, we have for any $\av$
\begin{eqnarray}\label{SimultaneousOutage}
 \onev(\gamma_k< R_k^{\star}(\av) )=  \onev(\alpha_k\gamma_1< \alpha_k R_1^{\star}(\av) ) = \onev(\gamma_1< R_1^{\star}(\av) ), \;\;\; k=2,\dots,K
\end{eqnarray}
where the first equality follows from (\ref{Prop}). The outage probability with the rate balancing scheme can be always written as
\begin{eqnarray*}
    P_{\rm out}^{\rm balance}(\gammav) &=& 1- \Pr( \cap_{k=1}^K \{ \gamma_k < R_k^{\star}(\av)\} ) \\
   &=& 1- \Pr(\gamma_k < R_k^{\star}(\av) )\Pr( \cap_{k'\neq k}
   \{\gamma_{k'} < R_{k'}^{\star}(\av)\}| \gamma_k < R_k^{\star}(\av) ) \\
    &=&1- \Pr(\gamma_k < R_k^{\star}(\av) )
\end{eqnarray*}
where the last equality follows since the equalities
(\ref{SimultaneousOutage}) imply $\Pr( \cap_{ {k'}\neq k}
\gamma_{k'} < R_{k'}^{\star}(\av)| \gamma_k < R_k^{\star}(\av) )=1$ for any $k$. This completes the second part. 

\subsection{Proof of Theorem~\ref{theorem:diversityDDS}} \label{proof:diversityDDS}
  In order to examine the diversity gain of the proposed scheduling scheme, we first remark
  \begin{align}
    \min_{ k \in \Ulb} \| \av^k \|^2
    &\geq \min_{ k \in \Ulb} \max_i |a_{ik}|^2\\ \label{eq:max-min}
    &\geq \max_i \min_{ k \in \Ulb} |a_{ik}|^2  \\
    &= \max_{\Uc\in\Pc_\Sc} \max_i  \min_{ k \in {\Uc}}|a_{ik}|^2
  \end{align}%
  where the (\ref{eq:max-min}) follows from the max-min inequality and the
  last equality holds since we can rewrite (\ref{eq:U_bar}) as
  \begin{equation}
    \Ulb = \arg_{\Uc} \max_{\Uc\in\Pc_\Sc} \max_{i=1\ldots B} \min_{k\in\Uc}  |a_{ik}|^2.
  \end{equation}
  by swapping the maximization over $\Uc$ and that over $i$. 
  To find the diversity order of the scheme, we need to look at the
  following near-zero behavior of the channel coefficients
  \begin{align}
    \Pr \left\{     \min_{ k \in \Ulb} \| \av^k
    \|^2    < \epsilon  \right\}  &\leq     \Pr \left\{
    \max_{\Uc\in\Pc_\Sc} \max_i \min_{ k \in {\Uc}} |a_{ik}|^2 <
    \epsilon \right\}\\
    &=   \left(  \Pr \left\{ \min_{ k \in {\Uc}} |a_{bk}|^2 <
    \epsilon \right\} \right)^{ B |\Pc_\Sc| } \label{eq:maxmax}\\
    &\leq   \left(  \sum_{k\in \Uc} \Pr \left\{  |a_{bk}|^2 <
    \epsilon \right\} \right)^{ B |\Pc_\Sc| } \label{eq:unionbound} \\
    &=   \left(  |\Uc| \Pr \left\{  |a_{bk}|^2 <
    \epsilon \right\} \right)^{ B |\Pc_\Sc| }\\
    &\doteq   \left(  |\Uc| \epsilon^{M+1-\Ks} \right)^{ B |\Pc_\Sc| }\\
    &\doteq   \epsilon^{B \frac{K}{\Ks} \left( M+1-\Ks \right)  },
  \end{align}%
  where (\ref{eq:maxmax}) follows from the fact that $\Uc$'s are disjoint in
  $\Pc_\Sc$ and that $\min_{ k \in {\Uc}} |a_{ik}|^2$ are
  independent for different $\Uc$ and $i$; (\ref{eq:unionbound}) is from
  the union bound. From lemma~\ref{lemma:diversity},
  (\ref{eq:diversity_scheduling}) is straightforward.
  For the upper bound of the diversity gain of the
  scheduling scheme (\ref{eq:Uub}), let us first write
  \begin{align}
    \max_{ \Uc \in \Qc } \min_{k\in\Uc} \|\av^k\|^2
    &\leq \max_{ \Uc \in \Qc } \|\av^{1}\|^2 \\
    &\leq B \max_i \max_{ \Uc \in \Qc } |a_{i1}|^2
  \end{align}%
  where the first inequality is from the fact that the worst user cannot
  be better than the first user; the second inequality is from
  $\|\av^1\|^2 \leq B \max_i |a_{i1}|^2$.
  From \cite[Theorem~1]{Jalden_selection}, we know that the diversity
  gain of $|a_{i1}|^2$ is $(K-\Ks+1)(M-\Ks+1)$. Therefore, it readily follows that the diversity gain of
    $\max_{ \Uc \in \Qc } \min_{k\in\Uc} \|\av^k\|^2 $ is
    upper-bounded by  $B(K-\Ks+1)(M-\Ks+1)$. 

\section*{Acknowledgment}
This work is partially supported by the European Commission through the FP7 project WiMAGIC (www.wimagic.eu) and
by the French cluster System@tic through the project POSEIDON. The
authors would like to thank Jakob Hoydis for his help in generating some numerical examples.

\bibliographystyle{IEEEtran}
\bibliography{downlink2}

\newpage
\begin{figure}
    \begin{center}
   \epsfxsize=5in
   \epsffile{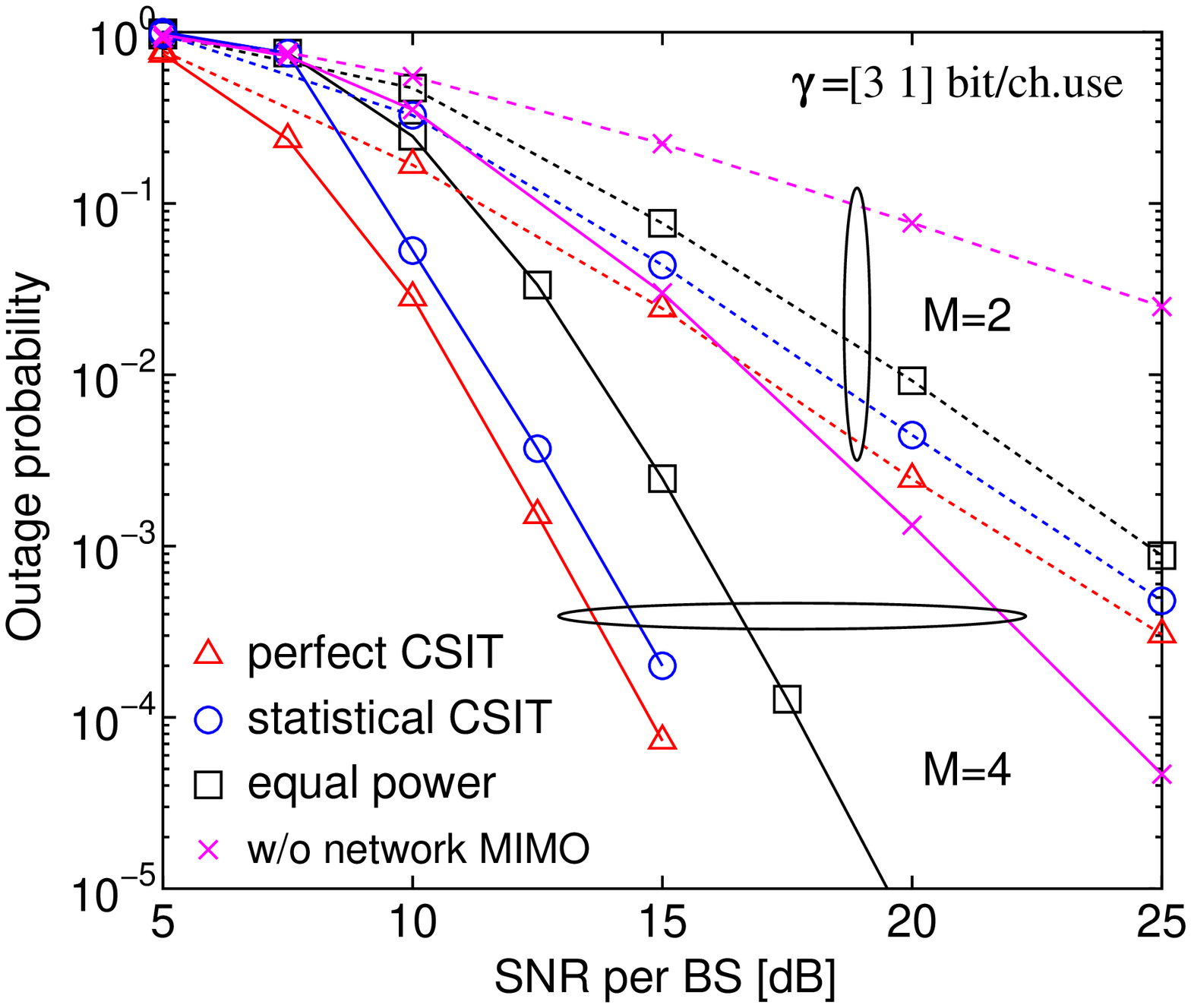}
    \end{center}
    \caption{Outage probability vs. SNR with $B=K=2$ and $M=2,4$.}
    \label{fig:AsymOutage}
\end{figure}

\begin{figure}
    \begin{center}
   \epsfxsize=5in
   \epsffile{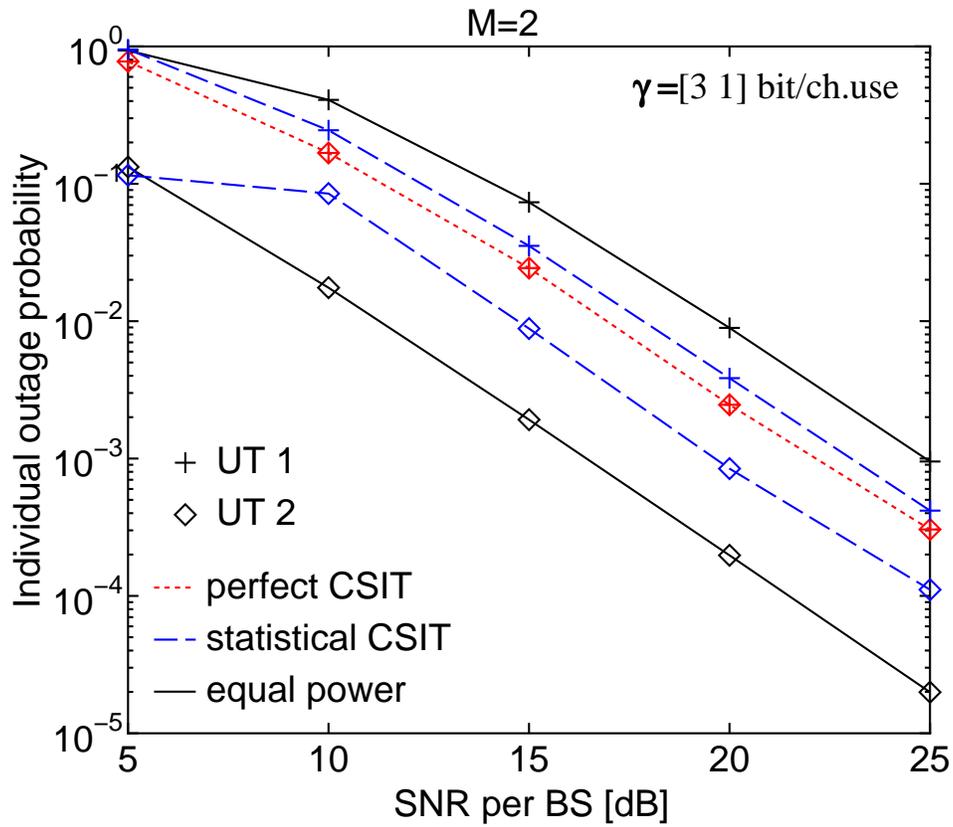}
    \end{center}
    \caption{Individual outage probabilities vs. SNR with $B=K=2$ and $M=2$.}
    \label{fig:IndividualOutage}
\end{figure}

\begin{figure}
    \begin{center}
   \epsfxsize=5in
   \epsffile{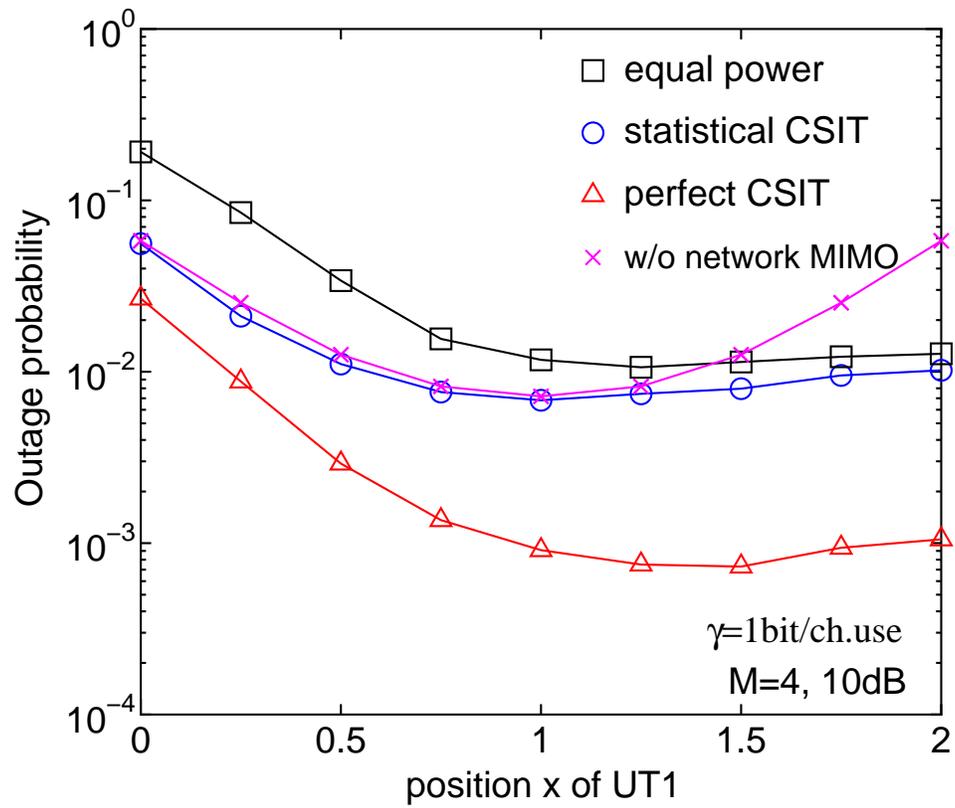}
    \end{center}
    \caption{Outage probability vs. location of UT1 with $B=K=2$ and $M=2,4$. }
 \label{fig:OutagevsDistance}
\end{figure}

\begin{figure}
    \begin{center}
   \epsfxsize=5in
   \epsffile{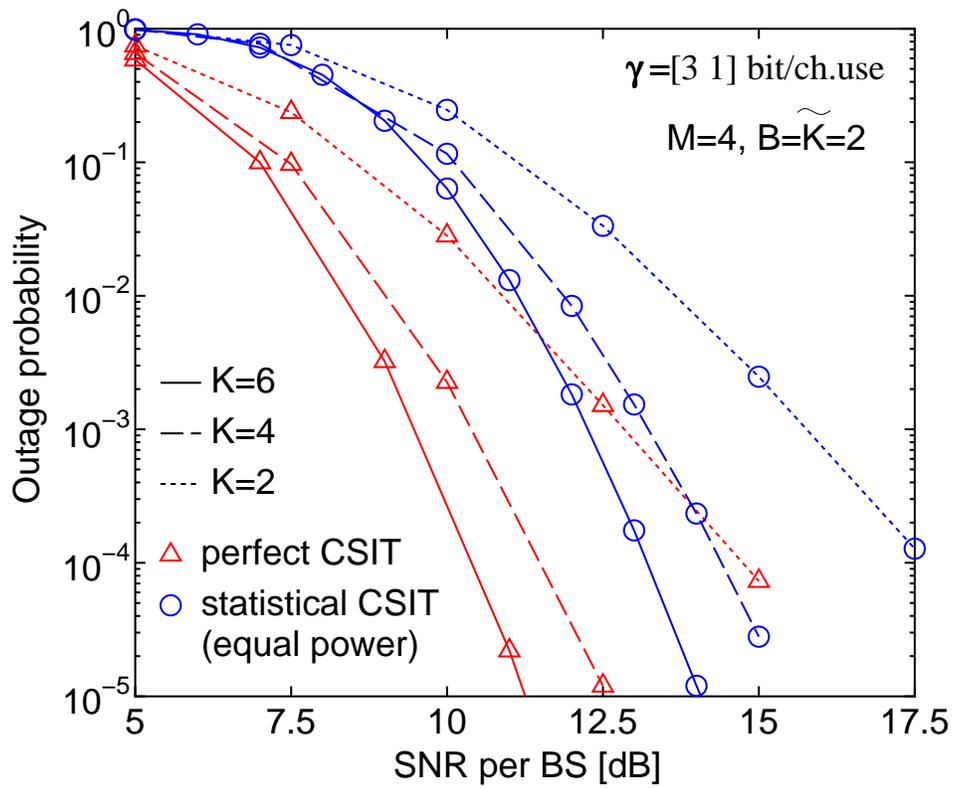}
    \end{center}
    \caption{Outage probability vs. SNR for many users with $B=\tilde{K}=2$ and $M=4$.}
 \label{fig:OutageManyUsers}
\end{figure}

\end{document}